\documentclass[twocolumn,aps,tightenlinefs,superscriptaddress,preprintnumbers]{revtex4}
\usepackage{mathrsfs}
\usepackage{rotating}
\usepackage{threeparttable}
\usepackage{lineno}

\usepackage{graphicx,color,dcolumn,booktabs,bm}
\usepackage{longtable,lscape}
\usepackage{amsmath}
\usepackage{indentfirst}
\usepackage{epsfig}
\usepackage{feynmf}   
\usepackage{epstopdf}   
\usepackage{slashed}  
\usepackage{cases}
\usepackage{color}
\usepackage{multirow}
\usepackage{ulem}
\usepackage{graphicx,color,dcolumn,booktabs,bm}

\usepackage{verbatim}
\usepackage[colorlinks,linkcolor=red,anchorcolor=green,citecolor=blue]{hyperref}
\usepackage{orcidlink}

\newcommand{\beq}{\begin{equation}}
	\newcommand{\eeq}{\end{equation}}
\newcommand{\bitm}{\begin{itemize}}
	\newcommand{\eitm}{\end{itemize}}

\newcommand{\tabincell}[2]{\begin{tabular}{@{}#1@{}}#2\end{tabular}} 

\begin{document}
	\hyphenpenalty=10000
	
	
	\title{\quad\\[0.1cm]\boldmath  Search for the semileptonic decays $\Xi_c^0 \to \Xi^0\ell^+\ell^-$ at Belle}

\noaffiliation

  \author{J.~X.~Cui\,\orcidlink{0000-0002-2398-3754}}
  \author{Y.~B.~Li\,\orcidlink{0000-0002-9909-2851}} 
  \author{C.~P.~Shen\,\orcidlink{0000-0002-9012-4618}} 
  \author{I.~Adachi\,\orcidlink{0000-0003-2287-0173}} 
\author{H.~Aihara\,\orcidlink{0000-0002-1907-5964}} 
\author{S.~Al~Said\,\orcidlink{0000-0002-4895-3869}} 
\author{D.~M.~Asner\,\orcidlink{0000-0002-1586-5790}} 
\author{T.~Aushev\,\orcidlink{0000-0002-6347-7055}} 
\author{R.~Ayad\,\orcidlink{0000-0003-3466-9290}} 
\author{V.~Babu\,\orcidlink{0000-0003-0419-6912}} 
% ...
\collaboration{The Belle Collaboration}
	
	\begin{abstract}
		Using the full data sample of 980 $\mathrm{fb}^{-1}$ collected with the Belle detector at the KEKB asymmetric energy electron-positron collider, we report the results of the first search for the rare semileptonic decays $\Xi_c^0 \to \Xi^0\ell^+\ell^-$ ($\ell=e$ or $\mu)$. No significant signals are observed in the $\Xi^0\ell^+\ell^-$ invariant-mass distributions. Taking the decay $\Xi_c^0 \to \Xi^- \pi^+$ as the normalization mode, we report 90\% credibility upper limits on the branching fraction ratios ${\cal{B}} (\Xi_c^0 \to \Xi^0 e^+ e^-) / {\cal{B}}(\Xi_c^0\to \Xi^-\pi^+) < 6.7 \times 10^{-3}$ and ${\cal{B}} (\Xi_c^0 \to \Xi^0 \mu^+ \mu^-) / {\cal{B}}(\Xi_c^0\to \Xi^-\pi^+) < 4.3 \times 10^{-3}$ based on the phase-space assumption for signal decays. The 90\% credibility upper limits on the absolute branching fractions of ${\cal{B}} (\Xi_c^0 \to \Xi^0 e^+ e^-)$ and ${\cal{B}} (\Xi_c^0 \to \Xi^0 \mu^+ \mu^-)$ are found to be $9.9 \times 10^{-5}$ and $6.5 \times 10^{-5}$, respectively.
		
	\end{abstract}
	
	
	\maketitle
	\section{\boldmath Introduction}

	In the Standard Model (SM), the weak-current interaction has an identical coupling to all lepton generations, which allows Lepton Flavor Universality (LFU) to be tested in the semileptonic decays of the hadrons.
	Theoretically, the study of semileptonic decays of baryons has complications that are not present in the study of analogous decays of mesons as the contributions from $W$-exchange transitions lead to sensitivity to the helicity structure of the effective Hamiltonian~\cite{Xic02xi0llTheoryPrediction,theory1,theory2,theory3}. 
	Furthermore, the hadronic form factors are not as well known for baryons as they are for mesons.
    Thus, the experimental results on baryonic semileptonic decays give important inputs for lattice quantum chromodynamics and other theoretical models.

	Experimentally, few baryonic neutrino-less semileptonic decays have been observed. 
	Of the light-baryon octet and bottom baryon decays, the branching fractions for $\Xi^0 \to \Lambda^0 e^+e^-$, $ \Sigma^+ \to p\mu^+\mu^-$, and $\Lambda_b \to \Lambda\mu^+\mu^-$ have been measured~\cite{Xi02LambdaeeNA482007,Sigmap2pmumuHyperCP2005,Lambdab02lambdamumuCDF2011,Lambdab02lambdamumuLHCb2013,Lambdab02lambdamumuLHCb2015}.
	However, no evidence has been found for similar decays of charmed baryons. 
	Among these decays, only $\Lambda_c^+ \to p\ell^+\ell^-$ ($\ell=e$ or $\mu$) decays were searched for.
	Upper limits on the branching fractions, at 90\% credibility, were first set by the BaBar collaboration at ${\cal{B}}(\Lambda_c^+ \to pe^+e^-) < 5.5 \times 10^{-6}$ and ${\cal{B}}(\Lambda_c^+ \to p\mu^+\mu^-) < 44 \times 10^{-6}$~\cite{Lambdacp2pllBARBAR2011}.
	LHCb then placed a much tighter limit, at the 90\% confidence level, on ${\cal{B}}(\Lambda_c^+ \to p\mu^+\mu^-)$ at $7.7\times10^{-8}$~\cite{Lambdacp2pmumuLHCb2018}.
	Particularly, the $\Lambda_c^+ \to p\ell^+\ell^-$ decays receive both single-quark transition via the Flavor Changing Neutral Current (FCNC) process and $W$-exchange contributions.

	The FCNC process is forbidden at tree level in the SM by the Glashow-Iliopoulos-Maiani mechanism~\cite{gem}. 
	However, some tensions have been reported recently in $B$ meson decays involving the $b\to s\ell^+\ell^-$ processes via LFU observables and angular analysis~\cite{RKplusLHCb2021,RKplusLHCb2019,RKplusLHCb2014,RKstar0LHCb2017,AngB02Kstar0LHCb2020,AngBplus2KstarplusLHCb2021}, whereas recently LHCb reported the disappearance of the anomaly on LFU~\cite{RKLHCb2023}.
	Hence, the study of semileptonic decays of baryons provides an opportunity to test the SM, and also can help in the understanding of the recent anomalies in meson FCNC processes.

	The lack of studies on semileptonic decays of charmed baryons provides a strong motivation for further research on these decays. 
	The $\Xi_c^0\to\Xi^0\ell^+\ell^-$ decays, which are related to the $W$-exchange contribution in $\Lambda_c^+ \to p\ell^+\ell^-$ decays under SU(3) flavor symmetry, have not been experimentally measured yet.
	Measurement of both $\Xi_c^0 \to \Xi^0 e^+e^-$ and $\Xi_c^0 \to \Xi^0 \mu^+\mu^-$ decay rates would also allow an LFU test to be performed. 
	Based on the SU(3) flavor symmetry and the recent experimental result on ${\cal{B}}(\Lambda_c^+ \to p\mu^+\mu^-)$~\cite{Lambdacp2pmumuLHCb2018}, the upper limits at the 68.3\% confidence level on the branching fractions of the Cabibbo-favored modes $\Xi_c^0 \to \Xi^0 \ell^+\ell^-$ are predicted to be ${\cal{B}}(\Xi_c^0 \to \Xi^0 e^+e^-)<2.35\times10^{-6}$ and ${\cal{B}}(\Xi_c^0 \to \Xi^0 \mu^+\mu^-)<2.25\times10^{-6}$~\cite{Xic02xi0llTheoryPrediction}.

	In this paper, we show the results of the first search for the $\Xi_c^0 \to \Xi^0\ell^+\ell^-$ decays using the full data sample of 980 $\mathrm{fb}^{-1}$ collected with the Belle detector~\cite{BELLEdetector}.
	The decay $\Xi_c^0\to \Xi^-\pi^+$ is used as the normalization mode.

	\section{\boldmath The data sample and the belle detector}
	This analysis is based on data recorded at or near the $\Upsilon(nS)$ ($n = 1 - 5$) resonances by the Belle detector~\cite{BELLEdetector} at the KEKB asymmetric energy electron-positron collider~\cite{KEKBcollider}.
	The Belle detector is a large solid-angle magnetic spectrometer consisting of a silicon vertex detector, a 50-layer central drift chamber (CDC), an array of aerogel threshold Cherenkov counters (ACC), a barrel-like arrangement of time-of-flight scintillation counters (TOF), an electromagnetic
	calorimeter (ECL) comprised of CsI(Tl) crystals located inside a superconducting solenoid coil that provides a $1.5~\hbox{T}$ magnetic field, and an iron flux return placed outside the coil, which is instrumented to detect $K^{0}_{L}$ mesons and to identify muons (KLM).
	
	Signal Monte Carlo (MC) events are generated using {\sc EVTGEN}~\cite{evtgen} to determine signal shapes, optimize the selection criteria, and calculate the reconstruction efficiencies.
	The generated $e^+e^- \to c\bar{c}$ events are simulated using {\sc pythia}~\cite{pythia} with a specific Belle configuration.
	The $\Xi_c^0$ particles in signal MC simulation decay to $ \Xi^0 e^+ e^-$, $\Xi^0 \mu^+ \mu^-$, and $\Xi^-\pi^+ $ using a phase-space model.
	These events are processed by a detector simulation based on {\sc geant3}~\cite{geant}.
	The Belle generic MC samples, which contain the MC samples of $\Upsilon(1S,2S,3S)$ decays, $\Upsilon(4S) \to B^{+}B^{-}/B^{0}\bar{B}^{0}$, $\Upsilon(5S) \to B_{s}^{(*)}\bar{B}_{s}^{(*)}$, and $e^+ e^- \to q\bar{q}$ $(q=u, d, s, c)$ at center-of-mass (c.m.)\ energies, $\sqrt{s}$, of 9.46, 10.024, 10.355, 10.52, 10.58, and 10.867~GeV with two times the total integrated luminosity of data, are used to study possible peaking backgrounds and verify the event selection criteria.

	\section{\boldmath Event selection criteria}
	For well-reconstructed charged tracks, except those from
	$\Xi^- \to \Lambda \pi^-$ and $\Lambda \to p \pi^-$ decays, the impact parameters perpendicular to and along the beam direction with respect to the nominal interaction point (IP) are required to be less than 0.1~cm and 2~cm, respectively. 
	Particle identification (PID) is applied to the reconstructed tracks. 
	Pions, kaons, and protons are distinguished based on specific ionisation in the CDC, time measurement in the TOF, and the response of the ACC: this information is combined to form a likelihood ${\mathcal L}_i$ for each particle hypothesis $i$, where $i$ = $\pi$, $K$, or $p$.
	Related likelihoods are used to identify leptons: electron identification also includes a comparison of track and ECL cluster	information, and muon identification is based on an extrapolation of the	particle track, and hits in the KLM~\cite{BellePID1,BellePID2,BellePID3}.

	The $\Lambda$ candidates are reconstructed via $\Lambda \to p \pi^-$ decay using a Kalman filter~\cite{b2fit} with fitted $\chi^2$ probability, $P_{\chi^2}$, greater than 0. The reconstructed mass should be within $\pm$3.5~MeV/$c^2$ of the nominal mass~\cite{pdg}, corresponding to approximately 2.5 times of the mass resolution ($\sigma$). The transverse distance for reconstructed $\Lambda$ vertex with respect to the IP is required to be greater than 0.35~cm. A loose PID requirement is applied on the proton with ${{\cal{L}}_p}/{({\cal{L}}_p + {\cal{L}}_K)}$ $>$ 0.2 and ${{\cal{L}}_p}/{({\cal{L}}_p + {\cal{L}}_{\pi})}$ $>$ 0.2. And $\cos(\alpha_{xyz}(\Lambda))$ is required to be larger than 0. 
	Hereinafter, $\alpha_{xyz}(i)$ is defined as the angle between the vector from the IP to the fitted decay vertex and the momentum vector of the reconstructed particle $i$; $\alpha_{xy}(i)$ is defined as the angle between the projections of these vectors on the plane perpendicular to the beam direction.

	Each $\pi^0$ candidate is reconstructed from a pair of photons with energy larger than 30 MeV in the barrel region of the ECL ($-0.63<\cos\theta<0.85$) or larger than 50 MeV in the endcaps ($-0.91<\cos\theta<-0.63$ or $0.85<\cos\theta<0.98$). Here, $\theta$ is the polar angle with respect to the detector axis, with the $\theta=0$ direction aligned approximately with the $e^-$ beam.
	The reconstructed invariant mass of the $\pi^0$ candidates is required to be within $\pm$17.4~MeV/$c^2$ ($\sim3\sigma$) of the $\pi^0$ nominal mass. 
	A mass-constrained fit is applied to the $\pi^0$ candidates, and the momenta of the fitted $\pi^0$ candidates in the laboratory frame are required to exceed 0.15 GeV/$c$. 
	
	The $\Xi^-\to\Lambda\pi^-$ decays are selected using the following criteria.
	The $\pi^-$ track is required to have a transverse momentum higher than 50~MeV/$c$.
	A TreeFit algorithm ~\cite{b2fit} which performs global decay chain vertex fitting for a particular process has been applied to the $\Xi^-$ decay chain with $P_{\chi^2} > $ 0 required. 
	The decay chain is required to satisfy $\cos(\alpha_{xyz}(\Xi^-))$ $>$ 0 and $\cos(\alpha_{xy}(\Lambda))/\cos(\alpha_{xy}(\Xi^-))$ $<$ 1.006. 
	The distances of the decay vertices of the reconstructed candidates from the IP, denoted as $L_i$, should satisfy $L_{\Lambda}>L_{\Xi^-}>0.1$ cm.		
	The reconstructed mass should be within $\pm$5~MeV/$c^2$ ($\sim2.5\sigma$) of the nominal mass~\cite{pdg}.

	The $\Xi^0$ is reconstructed by combining the selected $\Lambda$ and $\pi^0$ candidates.
	A TreeFit~\cite{b2fit,hyperons} to the $\Xi^0$ decay chain is applied with $P_{\chi^2} > $ 0 required.
	Since the $\pi^0$ from $\Xi^0$ decay has negligible vertex position information, the fit is performed with the following steps.
	Firstly, taking the IP as the point of origin of the $\Xi^0$, the point of intersection of the $\Xi^0$ trajectory and the reconstructed $\Lambda$ trajectory is	found.
	Then, this position is taken as the decay location of the $\Xi^0$ hyperon, and the $\pi^0$ is then re-made using this position as its point of origin.
	Only those combinations with the decay location of the $\Xi^0$ indicating a positive $\Xi^0$ path length are retained.
	The decay chain is also required to satisfy $\cos(\alpha_{xyz}(\Xi^0))$ $>$ 0, $\cos(\alpha_{xy}(\Xi^0)) > \cos(\alpha_{xy}(\Lambda))$, and $L_{\Lambda}>L_{\Xi^0}>0.5$ cm.
    The reconstructed mass should be within $\pm$12~MeV/$c^2$ ($\sim2.5\sigma$) of the nominal mass. 
	Backgrounds are studied using sideband samples: $\Xi^0$ candidates whose invariant mass differs by between 20 and 44 MeV/$c^2$ from the nominal value~\cite{pdg}.

	For the normalization channel $\Xi_c^0\to\Xi^-\pi^+$, the selected $\Xi^-$ hyperons are combined with selected $\pi^+$ candidates identified with ${{\cal{L}}_{\pi}}/({{\cal{L}}_{\pi} + {\cal{L}}_K})$ $>$ 0.2 and ${{\cal{L}}_{\pi}}/({{\cal{L}}_{\pi} + {\cal{L}}_p})$ $>$ 0.2. 
	To reconstruct the signal modes $\Xi_c^0 \to\Xi^0\ell^+\ell^-$, the $\Xi^0$ candidate is combined with a pair of lepton tracks, $e^+e^-$ or $\mu^+\mu^-$, which are identified with ${{\cal{L}}_{e}}/({{\cal{L}}_{e} + {\cal{L}}_{{\rm non}-e}})$ $>$ 0.9 and ${{\cal{L}}_{\mu}}/({{\cal{L}}_{\mu} + {\cal{L}}_K +{\cal{L}}_{\pi}})$ $>$ 0.9 for electrons and muons, respectively, where ${\cal{L}}_{{\rm non}-e}$ is the likelihood for non-electron tracks.
	The $\Xi_c^0$ candidates should be consistent with originating from the IP and pass the vertex and mass-constrained fits with $P_{\chi^2}$ $>$ 0.01 to the whole decay chain including the intermediate states, $\Xi^0$, $\Xi^-$, $\Lambda$, and $\pi^0$~\cite{b2fit}. 
	To reduce combinatorial background, especially those from $B$ meson decays, the scaled momentum for the $\Xi_c^0$ candidate, $x_p = p^{*}_{\Xi_c^0}c/\sqrt{s/4-M^{2}_{\Xi_c^0}c^{4}}$, is required to be greater than 0.5, where $p^{*}_{\Xi_c^0}$ is the momentum of $\Xi_c^0$ candidate in the $e^+e^-$ c.m. frame and $M_{\Xi_c^0}$ is the invariant mass of the $\Xi_c^0$ candidate.
	To suppress background from photon conversion for $\Xi_c^0 \to \Xi^0e^+e^-$ decay, the $e^+e^-$ pair is required to have invariant mass greater than 0.1 GeV/$c^2$. 
	Each of the electron candidates is also combined with every opposite-charged particle in the event, using the electron hypothesis: the invariant mass of all such pairs is required to be greater than 0.1 GeV/$c^2$.
	In events where there is at least one candidate, the average number of candidates is about 1.3.
	All candidates are retained.

	The selection criteria on the invariant mass of the electron pair, $P_{\chi^2}$ for the $\Xi_c^0$ decay chain, and scaled momentum $x_p$ in this analysis are optimized by maximizing the Punzi figure-of-merit, $\varepsilon/(3/2+\sqrt{B})$~\cite{PunziFOM}. 
	Here, `$3$' is the desired significance level, $\varepsilon$ is the detection efficiency of the $\Xi_c^0 \to \Xi^0e^+e^-$ mode based on signal MC simulation, and $B$ is the number of the normalized generic MC events in the signal range, $2.32< M_{\Xi^0e^+e^-}<2.50$~GeV/$c^2$ ($>$ 95\% signal events retained).
	These requirements are also found to be optimal for 
	$\Xi_c^0 \to \Xi^0 \mu^+\mu^-$, so they are applied for both channels.

	\section{\boldmath branching fraction measurement}
\begin{figure}[htbp]
		\begin{center}
			\includegraphics[width=7cm]{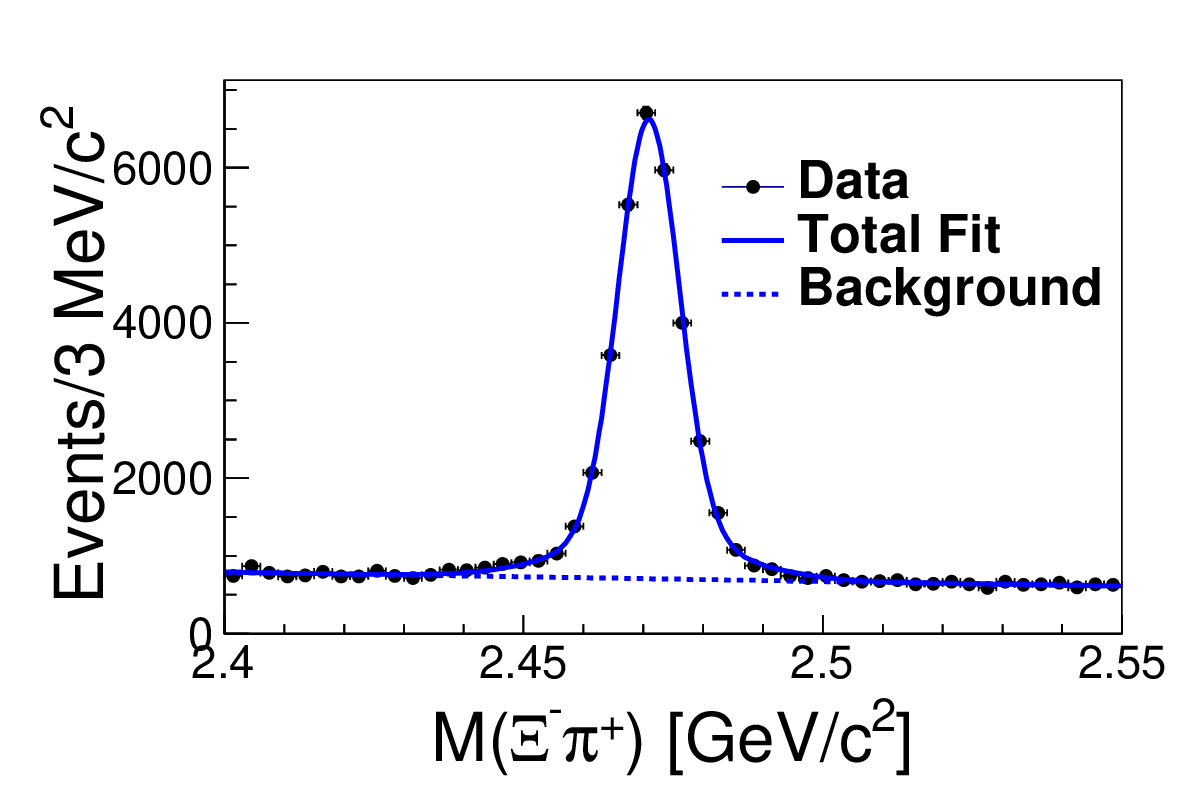}
			\caption{The invariant-mass distribution of $\Xi^-\pi^+$ combinations in data. 
				The dots with error bar represent the data,
				the solid curve shows the best-fit result, 
				and the blue dashed curve shows the fitted backgrounds.}
			\label{xic02xipidata}
		\end{center}
	\end{figure}

	For the reference mode, $\Xi_c^0 \to \Xi^- \pi^+$, the above selection criteria for $\Xi^-$ and $\Xi_c^0$ candidates are applied.
	Figure~\ref{xic02xipidata} shows the invariant-mass distribution of $\Xi^-\pi^+$ combinations from data, together with the result of an unbinned extended  maximum-likelihood (EML) fit. 
	In the fit, the signal shape of $\Xi_c^0$ candidates is parameterized by a double-Gaussian function, and the background shape is described by a first-order polynomial.
	The parameters are free in the fit. 
	The fitted signal yield is 28937$\pm$272.
	
		\begin{figure}[htbp]
		\begin{center}
			\includegraphics[width=7cm]{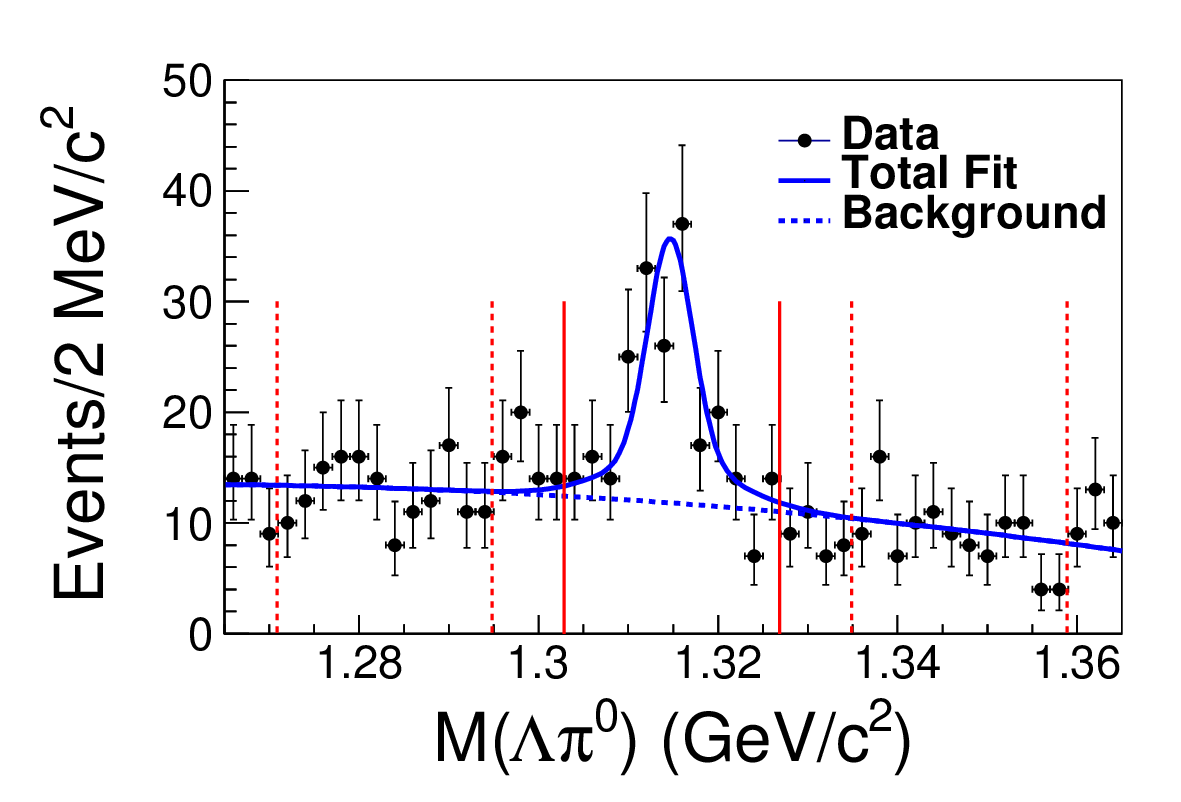}	
			\caption{The invariant-mass distributions of $\Xi^0\to\Lambda\pi^0$ candidates before combining with the $\ell^+\ell^-$ pairs in the selected $\Xi_c^0$ signal regions in the data.
				The dots with error bars represent the data, 
				the solid curve shows the total best-fit result, the dashed curve shows the background shape, and the solid and dashed lines show the signal and sideband regions of $\Xi^0$ candidates, respectively.}
			\label{mxi0data}
		\end{center}
	\end{figure}
	
	After applying the selection criteria introduced in the last section, 
	Fig.~\ref{mxi0data} shows the invariant-mass spectrum for the reconstructed $\Xi^0$ candidates before combining with the lepton pairs in the selected $\Xi_c^0$ signal regions from data, together with the fit result.
	Here, a double-Gaussian function is used to model the signal shape, and a second-order polynomial is used for the background. The signal shape parameters are fixed to the values found in signal MC, while the background parameters are free in the fit.

\begin{figure*}[htbp]
	\begin{center}
		\includegraphics[width=7cm]{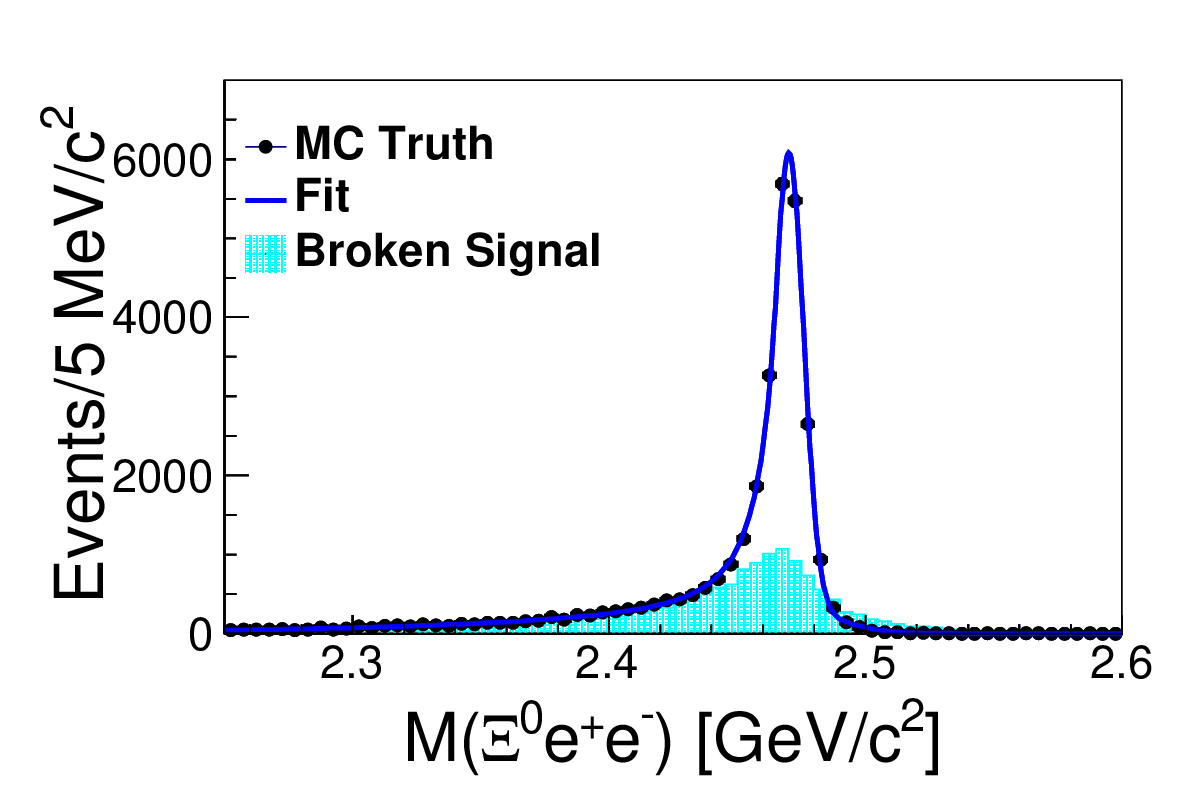}	 \put(-30,100){\bf (a)}
		\includegraphics[width=7cm]{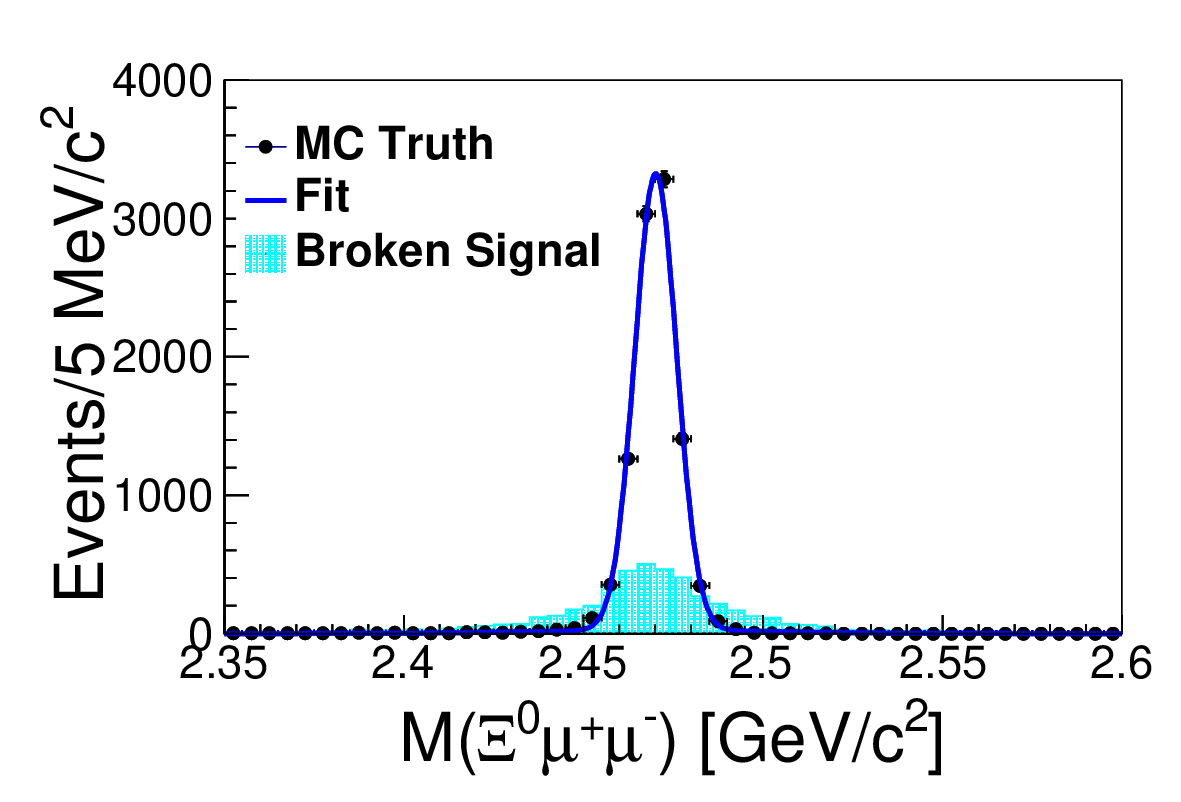}	\put(-30,100){\bf (b)}
		
		\caption{The invariant-mass distributions of (a) $\Xi^0e^+e^-$ and (b) $\Xi^0\mu^+\mu^-$ combinations in signal MC simulation. 
		Points with error bars show the correctly reconstructed signal, the blue solid curves show the results of the fit to the signal shape, and the cyan shaded histograms show the broken signal distributions.}
		\label{xic02xi0llmc}
	\end{center}
\end{figure*}

\begin{figure*}[htbp]
	\begin{center}
		\includegraphics[width=7cm]{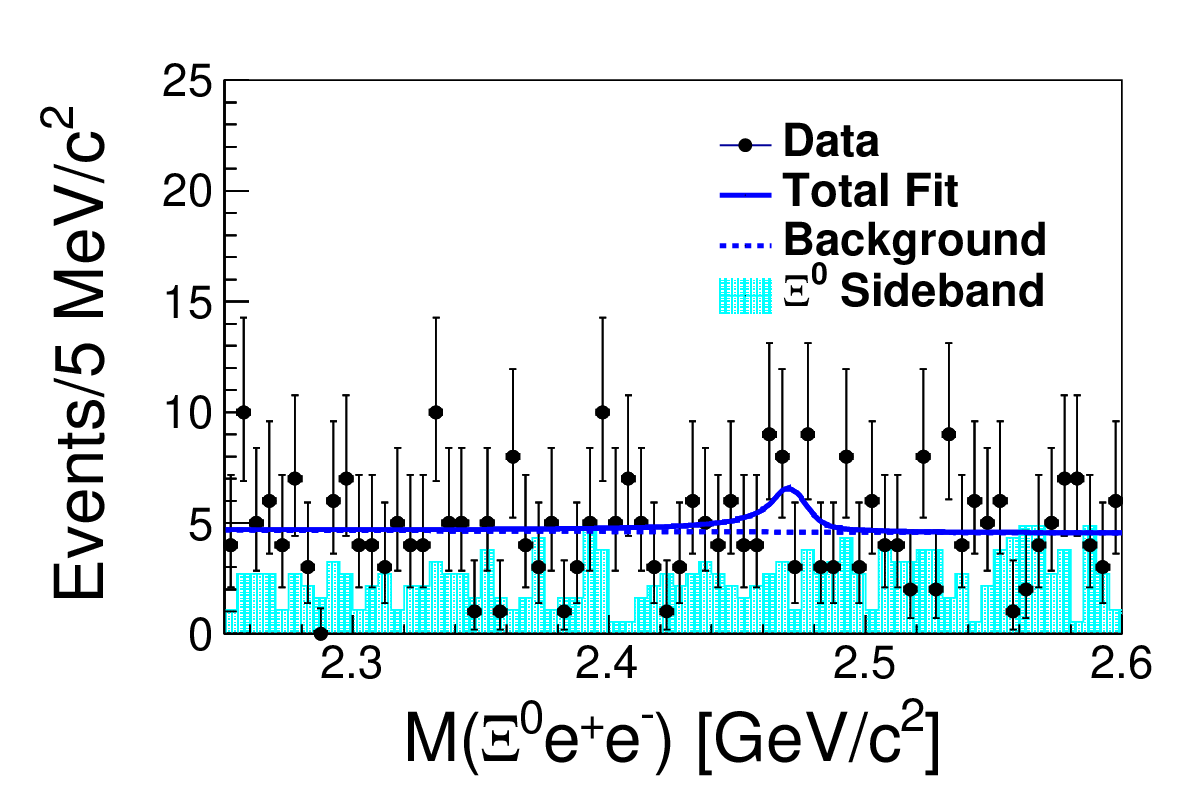} \put(-150,100){\bf (a)}
		\includegraphics[width=7cm]{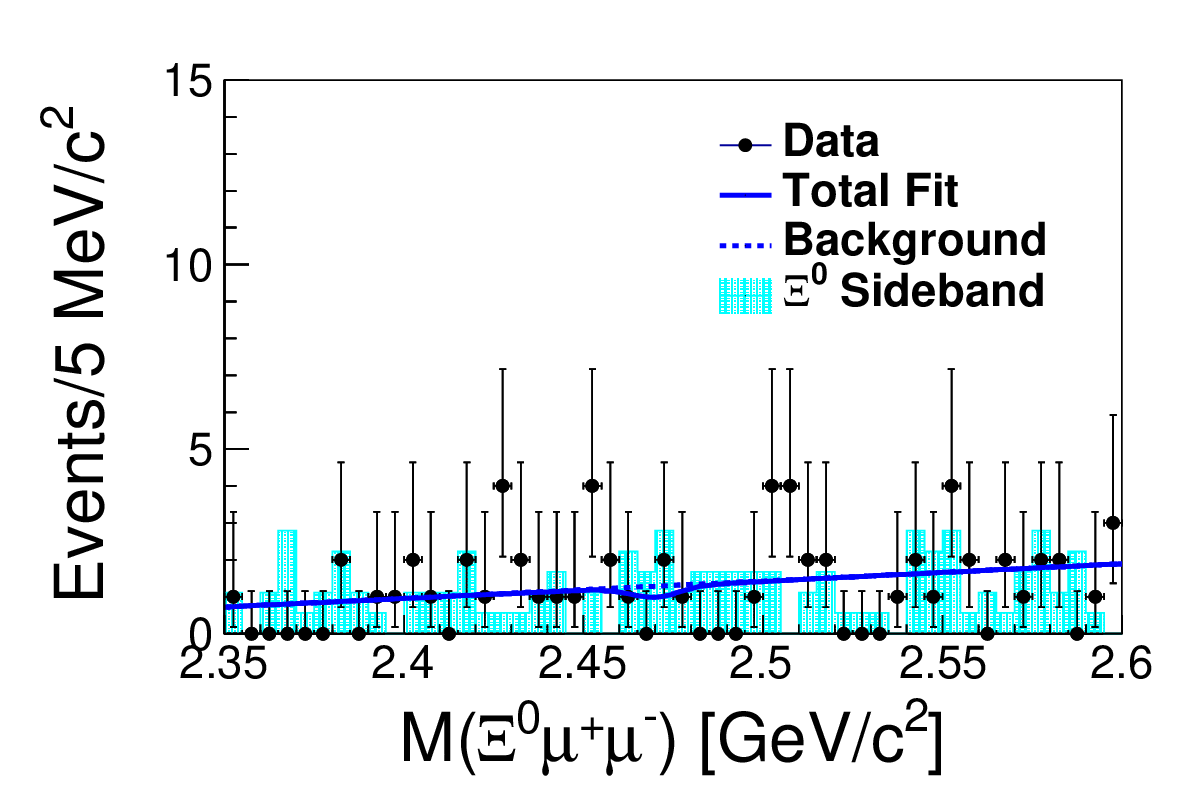}	\put(-150,100){\bf (b)}
		\caption{The invariant-mass distributions of (a) $\Xi^0e^+e^-$ and (b) $\Xi^0\mu^+\mu^-$ combinations in data. 
		Points with error bars show the data, the solid curves show the best-fit results, and the blue dashed curves show the background component in the fits.
		Cyan shaded histograms show the normalised $\Xi^0$ sidebands.}
		\label{xic02xi0lldata}
	\end{center}
\end{figure*}

	The invariant-mass distributions of $\Xi^0e^+e^-$ and $\Xi^0\mu^+\mu^-$ from signal MC simulations and data are shown in Fig.~\ref{xic02xi0llmc} and Fig.~\ref{xic02xi0lldata}, respectively, together with the unbinned EML fit results to the true signal distributions from signal MC events and spectra from data. 
	To take energy loss due to bremsstrahlung into account, the shapes of correctly reconstructed $\Xi_c^0$ candidates are described by two Crystal Ball functions~\cite{cbfunction} for the di-electron mode, while a double-Gaussian function is used as the signal shape for the di-muon mode. 
	Incorrectly reconstructed signal candidates (``broken signal'') have a broader distribution in signal MC simulation, shown by the cyan-shaded histograms in Fig.~\ref{xic02xi0llmc}.
	Broken signal is mainly due to incorrectly selected photons in $\Xi^0$ reconstruction.
	Similar to the treatment in Ref.~\cite{brokenSignal}, we extract the shape of the broken signal from MC simulation via rookeyspdf~\cite{rookeyspdf}, and treat it as a distinct component in the final $\Xi_c^0$ signal yield extraction.
	
	The peaking background components are determined using the algorithm of Ref.~\cite{topoana}, and we find them negligible.

	No significant $\Xi_c^0 \to \Xi^0 \ell^+\ell^-$ signals are observed in the data.
	The cyan shaded histograms in Fig.~\ref{xic02xi0lldata} indicate the normalized $\Xi^0$ sidebands. 
	For the fits to data, the signal shapes are taken from the fits to the signal MC samples above with all the parameters fixed. 
	Here, the width of the signal shape is multiplied by a correction factor $R_{\sigma}={\sigma_{\rm data}/\sigma_{\rm MC}}=1.12\pm0.06$, where $\sigma_{\rm data}$ and $\sigma_{\rm MC}$ are the fitted resolutions of $\Xi_c^0\to\Xi^-\pi^+$ shapes from data and MC simulation, respectively.
	The broken signal shape is taken from the MC simulation as described above, and the ratio of the broken signal to correctly reconstructed signal events, $R_{\rm broken/signal}$, is fixed at 0.50 (0.46) for the $\Xi^0e^+e^-(\Xi^0\mu^+\mu^-)$ mode according to MC simulation.
	Linear functions with free parameters are used for the smooth background shapes. 
	The fitted $\Xi_c^0$ signal yields are 9.1$\pm$7.1 with a significance of 1.4$\sigma$ and $-$0.9$\pm$2.1 for $\Xi_c^0 \to \Xi^0 e^+e^-$ and $\Xi_c^0\to\Xi^0\mu^+\mu^-$ decays, respectively.

	\begin{table}
	\caption{The summarized values for branching fraction measurements of $\Xi_c^0 \to \Xi^0 \ell^+ \ell^-$ decays.
		Here,
		$N^{\rm fit}$ is the fitted signal yield,
		$N^{\rm UL}$ is the 90\% credibility upper limit on the number of signal events from data before considering systematic uncertainties, 
		${  {\cal{B}}^{\rm UL}}/  { {\cal{B}}(\Xi_c^0 \to \Xi^-\pi^+) }$ and ${\cal{B}}^{\rm UL}$ are the 90\% credible upper limits on the relative and absolute branching fractions, respectively, for the $\Xi_c^0 \to \Xi^0 \ell^+ \ell^-$ decays with systematic uncertainties included, 
		and ${\cal{B}}(\Xi_c^0 \to \Xi^-\pi^+) = (1.43\pm0.32)\%$ is taken from the Particle Data Group~\cite{pdg}. 
	}
	\centering
	\vspace{0.2cm}
	\hspace{0.2cm}
	\label{values}
	\begin{tabular}{c c c  }
		\hline\hline
		Modes& 	$\Xi_c^0 \to \Xi^0 e^+ e^-$  & $\Xi_c^0 \to \Xi^0 \mu^+ \mu^-$ \\
		\hline
		Efficiency (\%) & 1.58 & 0.53 \\
		$N^{\rm fit}$ &  9.1$\pm$7.1 & $-$0.9$\pm$2.1 \\
		$N^{\rm UL} $  &  19.9 &  4.5  \\	
		${  {\cal{B}}^{\rm UL}}/  { {\cal{B}}(\Xi_c^0 \to \Xi^-\pi^+) }$  & 6.7$\times 10^{-3}$  & 4.3$\times 10^{-3}$ \\
		${\cal{B}}^{\rm UL}$ & 9.9$\times 10^{-5}$ & 6.5$\times 10^{-5}$ \\
		\hline\hline	 			
	\end{tabular}	
\end{table}

	Assuming the signal branching fraction has a uniform prior probability density function, the Bayesian upper limit at 90\% credibility on the number of signal events ($N^{\rm UL}$) is determined by solving the equation ${{\int_{0}^{N^{\rm UL}}}{\cal{L}}(x) dx} = 0.90 {\int_{0}^{\infty}} {\cal{L}}(x) dx$, where $x$ is the number of fitted signal events and ${\cal{L}}(x)$ is the likelihood function in the fit to data.
	The upper limits at 90\% credibility on the relative branching fractions are calculated by 
		\begin{equation}
		\begin{split}
			&{\cal{B}}^{\rm UL} (\Xi_c^0 \to \Xi^0 \ell^+ \ell^-) / {\cal{B}}(\Xi_c^0\to \Xi^-\pi^+) = \\ &\frac{N^{\rm UL}(\Xi_c^0 \to \Xi^0\ell^+\ell^-)}{N^{\rm obs}(\Xi_c^0 \to \Xi^-\pi^+)} \times \frac{\varepsilon({\Xi_c^0 \to \Xi^-\pi^+})}{\varepsilon({\Xi_c^0 \to \Xi^0\ell^+\ell^-})} \times \frac{   {\cal{B}}(\Xi^- \to \Lambda \pi^-)}{ {\cal{B}}(\Xi^0 \to \Lambda \pi^0)},
		\end{split}
	\end{equation}
	separately for $\ell=e$ and $\ell=\mu$. 
	Here, 
	$N^{\rm UL}(\Xi_c^0 \to \Xi^0\ell^+\ell^-)$ and $\varepsilon({\Xi_c^0 \to \Xi^0\ell^+\ell^-})$ are the upper limits on signal yield in data and the reconstruction efficiencies according to MC simulations, respectively, of $\Xi_c^0 \to \Xi^0 \ell^+\ell^-$ decays, $N^{\rm obs}({\Xi_c^0 \to \Xi^-\pi^+})$ and $\varepsilon({{\Xi_c^0 \to \Xi^-\pi^+}})$ are the number of observed events in data and the reconstruction efficiency, respectively, of $\Xi_c^0 \to \Xi^-\pi^+$ decay, and the branching fractions are taken as ${\cal{B}}(\Xi_c^0 \to \Xi^-\pi^+) = (1.43\pm0.32)\%$,  ${\cal{B}}(\Xi^0 \to \Lambda \pi^0) = (99.524\pm0.012)\%$, and ${\cal{B}} (\Xi^- \to \Lambda \pi^-) = (99.887\pm0.035)$\% ~\cite{pdg}. 
	To take into account the systematic uncertainties detailed in the next section, the likelihood curve is convolved with a Gaussian function whose width equals the corresponding total multiplicative systematic uncertainty.
	The calculated 90\% credible upper limits on the numbers of signal events, and relative and absolute branching fractions in data, are summarized in Table~\ref{values}.
	The muon identification criterion used in this analysis effectively excludes tracks with a momentum too low to reach the KLM~\cite{BellePID3}: this leads to a reconstruction efficiency in the di-muon channel that is a factor of three lower than in the di-electron channel.

	\section{\boldmath Systematic Uncertainties}
	The systematic uncertainties in the measurements of the branching fractions are divided into two categories: multiplicative and additive systematic uncertainties.
	
	The sources of multiplicative systematic uncertainties include detection-efficiency-related uncertainties, branching fractions of intermediate states, and the fitting uncertainty for the normalization mode $\Xi_c^0 \to \Xi^- \pi^+$.
	The additive systematic uncertainties are the uncertainties in the fits to extract signal yields for $\Xi_c^0 \to \Xi^0 \ell^+\ell^-$ decays.
	
	The detection-efficiency-related uncertainties 
	include those for tracking efficiency, PID efficiency, $\pi^0$ and $\Lambda$ selection efficiencies, and are estimated based on the simulated MC samples. 
	Since there are four charged tracks in the final states for both signal and reference decay modes,	
	the uncertainty in tracking efficiency cancels in this analysis. 
	The proton PID uncertainties are found to be 0.6\% and 1.1\% for $\Xi_c^0 \to \Xi^0e^+e^-$ and $\Xi_c^0 \to \Xi^0\mu^+\mu^-$ modes, respectively, by taking into account the proton momentum differences with the normalization mode.
	Since the $\Lambda\to p\pi^-$ decay is reconstructed in each decay mode and no PID requirement is assigned for the pion track decay from $\Lambda$, the other sources of $\Lambda$ selection uncertainties cancel.
	Using the control samples of $D^{*+} \to D^0\pi^+$ with $D^0 \to K^-\pi^+$, the PID uncertainties are estimated to be 0.5\% and 0.6\% for pions from $\Xi_c^0$ and $\Xi^-$ decays, respectively, and are added linearly to be 1.1\% for the pion tracks in $\Xi_c^0 \to \Xi^-\pi^+$ decay.
	Based on the study of $J/\psi \to \ell^+\ell^-$ decay, the uncertainties from lepton identification are determined to be 3.2\% for electrons and 5.2\% for muons.
	The PID uncertainties here are summed in quadrature for different decay modes, assuming that those uncertainties are independent for $\Xi_c^0\to \Xi^0\ell^+\ell^-$ and $\Xi_c^0 \to \Xi^-\pi^+$ decays.
	The total PID systematic uncertainties for $\Xi_c^0 \to \Xi^0e^+e^-$ and $\Xi_c^0 \to \Xi^0\mu^+\mu^-$ decays are determined to be 3.5\% and 5.5\%, respectively.
	The systematic uncertainties for momentum-weighted $\pi^0$ selection efficiency are estimated to be 3.3\% and 3.0\% for $\Xi_c^0 \to \Xi^0e^+e^-$ and $\Xi_c^0 \to \Xi^0\mu^+\mu^-$ decays, respectively, according to a study of a $\tau^- \to \pi^-\pi^0\nu_{\tau}$ control sample. 
	Assuming these uncertainties to be uncorrelated, the uncertainties from PID and $\pi^0$ efficiencies are added in quadrature to yield the total multiplicative systematic uncertainties.
	
	For the measurements of ratios of branching fractions ${\cal{B}}^{\rm UL} (\Xi_c^0 \to \Xi^0 \ell^+ \ell^-) / {\cal{B}}(\Xi_c^0\to \Xi^-\pi^+)$, the uncertainties associated with branching fractions of intermediate states ${{\cal{B}}(\Xi^- \to \Lambda\pi^-)}$ and ${{\cal{B}}(\Xi^0 \to \Lambda\pi^0)}$ are 0.035\% and 0.012\%~\cite{pdg}, respectively, which are negligible. 
	In the measurements of absolute branching fractions, the uncertainty on ${\cal{B}}(\Xi_c^0 \to \Xi^-\pi^+)$ is 22.4\%~\cite{pdg}, which is the dominant contribution.
	
	In the fit to the $M(\Xi^-\pi^+)$ distribution from data for $\Xi_c^0 \to \Xi^-\pi^+$ decay, we change the fit range by $\pm$10\% and the order of the polynomial for the background shape, and the relative differences of the fitted signal yields are taken as the uncertainties.
	These uncertainties are added in quadrature: the total is 0.7\%.
	
	Additive systematic uncertainties due to the $\Xi^0\ell^+\ell^-$ invariant-mass fits are considered by re-performing the fits with all combinations of the following options:
	(1) change the resolution scale factor $R_{\sigma}$ by $\pm1\sigma$ of its uncertainty; (2) change the fit range by $\pm$10\%, (3) change the polynomial describing the background shape from first-order to second-order; and (4) multiply the fixed $R_{\rm broken/signal}$ ratios by the correction factors of 1.43 and 0.93 for wrong $\Xi^0$ combinations in di-electron and di-muon modes, respectively, which are calculated according to the ratios of the number of events in $\Xi^0$ sidebands from data over that from generic MC samples.
	
	For the measurements of the upper limits at 90\% credibility on the relative and absolute branching fractions of $\Xi_c^0 \to \Xi^0\ell^+\ell^-$ decays, the systematic uncertainties are taken into account in two steps.
	First, based on the study of the additive systematic uncertainties, the most conservative upper limits at 90\% credibility on the numbers of $\Xi_c^0$ signal events are 25.6 and 4.6 for di-election and di-muon modes, respectively.
	Then, the likelihood function of the case with the most conservative upper limit is convolved with a Gaussian function whose width equals the corresponding total multiplicative systematic uncertainty for each $\Xi_c^0 \to \Xi^0\ell^+\ell^-$ decay to get the final upper limit. 
	The multiplicative systematic uncertainties from different sources are summarized in Table~\ref{multSyst}.

	\linespread{1.2}
	\begin{table*}[htbp]
		\caption{The multiplicative systematic uncertainties (\%) on the measurements of relative and absolute branching fractions.}
		\vspace{0.2cm}
		\label{multSyst}
		\begin{tabular}{l c c }
			\hline\hline
			Source & $\Xi_c^0 \to \Xi^0e^+e^-$ & $\Xi_c^0 \to \Xi^0\mu^+\mu^-$ \\
			\hline
		Particle ID & 3.5 &5.5 \\
$\pi^0$ selection & 3.3 & 3.0 \\
${\cal{B}}(\Xi_c^0 \to \Xi^-\pi^+)$ & 22.4 & 22.4 \\ 
Fit of reference mode & 0.7 & 0.7 \\
\hline
Total [${\cal{B}} (\Xi_c^0 \to \Xi^0 \ell^+ \ell^-) / {\cal{B}}(\Xi_c^0\to \Xi^-\pi^+)$] & 4.9 & 6.3 \\
\hline
Total [${\cal{B}} (\Xi_c^0 \to \Xi^0 \ell^+ \ell^-)$] & 23.0 & 23.3 \\			
			
			\hline\hline
		\end{tabular}
	\end{table*}

	In this analysis, the simulated $\Xi_c^0 \to \Xi^0 \ell^+ \ell^-$ decays are generated by the phase space model, since the exact physics models for the decays are unknown, and no significant signals are observed in data. Thus, no systematic uncertainty due to the choice of the decay model is included. Instead, we provide the reconstruction efficiencies in ($M^{2}_{\ell^+\ell^-}$, $M^{2}_{\Xi^0 \ell^+}$) bins, which are shown in Table~\ref{eff1} and Table~\ref{eff2} for the di-electron and di-muon modes respectively.

	\begin{table*}
		\caption{The detection efficiencies (\%) in ($M^{2}_{e^+e^-}$, $M^{2}_{\Xi^0 e^+}$) bins for the $\Xi_c^0 \to \Xi^0e^+e^-$ decay mode.}
		\centering
		\vspace{0.2cm}
		\hspace{0.2cm}
		\label{eff1}
		\setlength{\tabcolsep}{1.2mm}{
			\begin{tabular}{c c c c c c c c }
				\hline\hline
				&  \multicolumn{7}{c}{  $ M^{2}_{\Xi^{0}e^{+}} $  ( GeV$^{2}/c^4 $ )} \\
				\hline
				\tabincell{c}{$ M^{2}_{e^{+}e^{-}}$  \\  (GeV$^{2}$/c$^4$)}  & [1.4, 2.1]  & [2.1, 2.8]  & [2.8, 3.5]  & [3.5, 4.2]  & [4.2, 4.9]  & [4.9, 5.6]  & [5.6, 6.3]    \\
				\hline
				
{[0,0.15]} 	   &  0.39 $ \pm $ 0.01  &  0.98 $ \pm $ 0.01  &  1.53 $ \pm $ 0.01  &  1.69 $ \pm $ 0.01  &  1.57 $ \pm $ 0.01  &  1.03 $ \pm $ 0.01  &  0.39 $ \pm $ 0.01  \\
{[0.15,0.3]} 	   &  0.65 $ \pm $ 0.01  &  1.20 $ \pm $ 0.01  &  1.65 $ \pm $ 0.01  &  1.78 $ \pm $ 0.01  &  1.63 $ \pm $ 0.01  &  1.08 $ \pm $ 0.01  &  0.51 $ \pm $ 0.01  \\
{[0.3,0.45]} 	   &  0.92 $ \pm $ 0.01  &  1.32 $ \pm $ 0.01  &  1.68 $ \pm $ 0.01  &  1.78 $ \pm $ 0.01  &  1.60 $ \pm $ 0.01  &  1.06 $ \pm $ 0.01  &  0.90 $ \pm $ 0.03  \\
{[0.45,0.6]} 	   &  1.19 $ \pm $ 0.01  &  1.48 $ \pm $ 0.01  &  1.77 $ \pm $ 0.01  &  1.83 $ \pm $ 0.01  &  1.62 $ \pm $ 0.01  &  1.15 $ \pm $ 0.01  &  ...                \\
{[0.6,0.75]} 	   &  1.30 $ \pm $ 0.02  &  1.61 $ \pm $ 0.01  &  1.87 $ \pm $ 0.01  &  1.91 $ \pm $ 0.01  &  1.63 $ \pm $ 0.01  &  1.17 $ \pm $ 0.01  &  ...                \\
{[0.75,0.9]} 	   &  ...                &  1.72 $ \pm $ 0.01  &  1.97 $ \pm $ 0.01  &  1.96 $ \pm $ 0.01  &  1.56 $ \pm $ 0.01  &  1.80 $ \pm $ 0.07  &  ...                \\
{[0.9,1.05]} 	   &  ...                &  1.80 $ \pm $ 0.01  &  2.07 $ \pm $ 0.01  &  1.95 $ \pm $ 0.01  &  1.53 $ \pm $ 0.01  &  ...                &  ...                \\
{[1.05,1.2]} 	   &  ...                &  1.75 $ \pm $ 0.01  &  2.04 $ \pm $ 0.01  &  1.81 $ \pm $ 0.01  &  1.42 $ \pm $ 0.02  &  ...                &  ...                \\
{[1.2,1.35]} 	   &  ...                &  1.60 $ \pm $ 0.02  &  1.78 $ \pm $ 0.01  &  1.67 $ \pm $ 0.01  &  ...                &  ...                &  ...                \\

				\hline\hline		
			\end{tabular}
		}
	\end{table*}

	\begin{table*}
		\caption{The detection efficiencies ($\times 10^{-3}$) in ($M^{2}_{\mu^+\mu^-}$, $M^{2}_{\Xi^0 \mu^+}$) bins for the $\Xi_c^0 \to \Xi^0\mu^+\mu^-$ decay mode.}
		\vspace{0.2cm}
		\hspace{0.2cm}
		\label{eff2}
		\setlength{\tabcolsep}{1.5mm}{
			\begin{tabular}{c c c c c c }
				\hline\hline
				
				&  \multicolumn{5}{c}{ $ M^{2}_{\Xi^{0}\mu^{+}} $  ( GeV$^{2}/c^4 $)} \\
				\hline
				\tabincell{c}{$ M^{2}_{\mu^{+}\mu^{-}} $ \\ (GeV$^{2}$/c$^4$) }   & [2.1, 2.8]  & [2.8, 3.5]  & [3.5, 4.2]  & [4.2, 4.9]  & [4.9, 5.6]    \\
				
				\hline
				
{[0,0.15]} 	   &  0.33 $ \pm $ 0.01  &  5.87 $ \pm $ 0.04  &  11.15 $ \pm $ 0.05  &  6.66 $ \pm $ 0.04  &  0.46 $ \pm $ 0.01  \\
{[0.15,0.3]} 	   &  0.56 $ \pm $ 0.01  &  6.68 $ \pm $ 0.03  &  9.99 $ \pm $ 0.04  &  6.01 $ \pm $ 0.03  &  0.35 $ \pm $ 0.01  \\
{[0.3,0.45]} 	   &  1.19 $ \pm $ 0.01  &  7.05 $ \pm $ 0.03  &  9.03 $ \pm $ 0.04  &  5.38 $ \pm $ 0.03  &  0.36 $ \pm $ 0.01  \\
{[0.45,0.6]} 	   &  1.75 $ \pm $ 0.02  &  7.05 $ \pm $ 0.03  &  8.30 $ \pm $ 0.03  &  4.60 $ \pm $ 0.03  &  0.38 $ \pm $ 0.01  \\
{[0.6,0.75]} 	   &  2.33 $ \pm $ 0.02  &  7.11 $ \pm $ 0.03  &  7.69 $ \pm $ 0.03  &  3.82 $ \pm $ 0.02  &  0.38 $ \pm $ 0.02  \\
{[0.75,0.9]} 	   &  3.02 $ \pm $ 0.02  &  6.92 $ \pm $ 0.03  &  7.20 $ \pm $ 0.03  &  3.38 $ \pm $ 0.02  &  1.55 $ \pm $ 0.40  \\
{[0.9,1.05]} 	   &  3.90 $ \pm $ 0.03  &  6.84 $ \pm $ 0.03  &  6.67 $ \pm $ 0.03  &  3.60 $ \pm $ 0.03  &  ...                \\
{[1.05,1.2]} 	   &  5.24 $ \pm $ 0.05  &  7.20 $ \pm $ 0.03  &  6.59 $ \pm $ 0.03  &  4.26 $ \pm $ 0.11  &  ...                \\
{[1.2,1.35]} 	   &  6.59 $ \pm $ 0.16  &  7.87 $ \pm $ 0.04  &  7.56 $ \pm $ 0.06  &  ...                &  ...                \\

				\hline\hline
			\end{tabular}
		}
	\end{table*}

	\section{\boldmath conclusion}
	In summary, using the entire data sample of 980 $\mathrm{fb}^{-1}$ collected with the Belle detector, we searched for the semileptonic decays $\Xi_c^0 \to \Xi^0\ell^+\ell^- $. 
	No significant signals are observed in the $\Xi^0\ell^+\ell^-$ invariant-mass distributions. 
	We determine 90\% credible upper limits on the relative branching fraction ratios ${\cal{B}} (\Xi_c^0 \to \Xi^0 e^+ e^-) / {\cal{B}}(\Xi_c^0\to \Xi^-\pi^+) < 6.7 \times 10^{-3}$ and ${\cal{B}} (\Xi_c^0 \to \Xi^0 \mu^+ \mu^-) / {\cal{B}}(\Xi_c^0\to \Xi^-\pi^+) < 4.3 \times 10^{-3}$.
	Taking ${\cal{B}}(\Xi_c^0 \to \Xi^-\pi^+) = (1.43\pm0.32)\%$~\cite{pdg}, 90\% credible upper limits on the absolute branching fractions ${\cal{B}} (\Xi_c^0 \to \Xi^0 e^+ e^-)$ and ${\cal{B}} (\Xi_c^0 \to \Xi^0 \mu^+ \mu^-)$ are determined to be $9.9 \times 10^{-5}$ and $6.5 \times 10^{-5}$ respectively. 
	
	Comparing with the theoretical predictions of the upper limits on the branching fractions of $\Xi_c^0 \to \Xi^0 \ell^+\ell^- $ decays, ${\cal{B}}(\Xi_c^0 \to \Xi^0 e^+e^-) < 2.35\times10^{-6}$ and ${\cal{B}}(\Xi_c^0 \to \Xi^0 \mu^+\mu^-) < 2.25\times10^{-6}$~\cite{Xic02xi0llTheoryPrediction}, 
	the experimental upper limits reported in this paper using a uniform-phase-space distribution are higher by an order-or-magnitude
	than those calculated using theoretical arguments and input from other experimental results.
	A more precise analysis based on larger data samples collected by Belle II is expected in the future.
	\section{\boldmath ACKNOWLEDGMENTS}
This work, based on data collected using the Belle detector, which was
operated until June 2010, was supported by 
the Ministry of Education, Culture, Sports, Science, and
Technology (MEXT) of Japan, the Japan Society for the 
Promotion of Science (JSPS), and the Tau-Lepton Physics 
Research Center of Nagoya University; 
the Australian Research Council including grants
DP210101900, 
DP210102831, 
DE220100462, 
LE210100098, 
LE230100085; 
Austrian Federal Ministry of Education, Science and Research (FWF) and
FWF Austrian Science Fund No.~P~31361-N36;
National Key R\&D Program of China under Contract No.~2022YFA1601903,
National Natural Science Foundation of China and research grants
No.~11575017,
No.~11761141009, 
No.~11705209, 
No.~11975076, 
No.~12135005, 
No.~12150004, 
No.~12161141008, 
and
No.~12175041, 
and Shandong Provincial Natural Science Foundation Project ZR2022JQ02;
the Czech Science Foundation Grant No. 22-18469S;
Horizon 2020 ERC Advanced Grant No.~884719 and ERC Starting Grant No.~947006 ``InterLeptons'' (European Union);
the Carl Zeiss Foundation, the Deutsche Forschungsgemeinschaft, the
Excellence Cluster Universe, and the VolkswagenStiftung;
the Department of Atomic Energy (Project Identification No. RTI 4002), the Department of Science and Technology of India,
and the UPES (India) SEED finding programs Nos. UPES/R\&D-SEED-INFRA/17052023/01 and UPES/R\&D-SOE/20062022/06; 
the Istituto Nazionale di Fisica Nucleare of Italy; 
National Research Foundation (NRF) of Korea Grant
Nos.~2016R1\-D1A1B\-02012900, 2018R1\-A2B\-3003643,
2018R1\-A6A1A\-06024970, RS\-2022\-00197659,
2019R1\-I1A3A\-01058933, 2021R1\-A6A1A\-03043957,
2021R1\-F1A\-1060423, 2021R1\-F1A\-1064008, 2022R1\-A2C\-1003993;
Radiation Science Research Institute, Foreign Large-size Research Facility Application Supporting project, the Global Science Experimental Data Hub Center of the Korea Institute of Science and Technology Information and KREONET/GLORIAD;
the Polish Ministry of Science and Higher Education and 
the National Science Center;
the Ministry of Science and Higher Education of the Russian Federation, Agreement 14.W03.31.0026, 
and the HSE University Basic Research Program, Moscow; 
University of Tabuk research grants
S-1440-0321, S-0256-1438, and S-0280-1439 (Saudi Arabia);
the Slovenian Research Agency Grant Nos. J1-9124 and P1-0135;
Ikerbasque, Basque Foundation for Science, and the State Agency for Research
of the Spanish Ministry of Science and Innovation through Grant No. PID2022-136510NB-C33 (Spain);
the Swiss National Science Foundation; 
the Ministry of Education and the National Science and Technology Council of Taiwan;
and the United States Department of Energy and the National Science Foundation.
These acknowledgements are not to be interpreted as an endorsement of any
statement made by any of our institutes, funding agencies, governments, or
their representatives.
We thank the KEKB group for the excellent operation of the
accelerator; the KEK cryogenics group for the efficient
operation of the solenoid; and the KEK computer group and the Pacific Northwest National
Laboratory (PNNL) Environmental Molecular Sciences Laboratory (EMSL)
computing group for strong computing support; and the National
Institute of Informatics, and Science Information NETwork 6 (SINET6) for
valuable network support.

\end{document}